# Ultimate charge transport regimes in doping-controlled graphene laminates: phonon-assisted processes revealed by the linear magnetoresistance.

*Mohsen Moazzami Gudarzi [1], Sergey Slizovskiy [1], Boyang Mao[1,3], Endre Tovari [4], Gergo Pinter [2], David Sanderson [2], Maryana Asaad [2], Ying Xiang [2], Zhiyuan Wang [2], Jianqiang Guo [2], Ben F. Spencer [2], Alexandra A. Geim [5], Vladimir I. Fal'ko [1,6,7] and Andrey V. Kretinin [1,2,6]\**

\* E-mail: andrey.kretinin@manchester.ac.uk

[1] Department of Physics and Astronomy, The University of Manchester, Oxford Road, Manchester, M13 9PL, UK

[2] Department of Materials, The University of Manchester, Oxford Road, Manchester, M13 9PL, UK

[3] Cambridge Graphene Centre, Department of Engineering, University of Cambridge, 9 JJ Thomson Ave, Cambridge CB3 0FA

[4] Department of Physics, Institute of Physics, Budapest University of Technology and Economics, Műegyetem rkp. 3, H-1111 Budapest, Hungary

[5] Department of Physics, Harvard University, Cambridge, MA, USA

[6] National Graphgene Institute, The University of Manchester, Oxford Road, Manchester, M13 9PL, UK

[7] Henry Royce Institute for Advanced Materials, The University of Manchester, Oxford Road, Manchester, M13 9PL, UK

ABSTRACT. Understanding and controlling the electrical properties of solution-processed 2D materials is key to further printed electronics progress. Here we demonstrate that the thermolysis of the aromatic intercalants utilized in nanosheet exfoliation for graphene laminates opens the route to achieving high intrinsic mobility and simultaneously controlling doping type (*n*- and *p*-) and concentration over a wide range. We establish that the intra-flake mobility is high by observing a linear magnetoresistance of such solution-processed graphene laminates and using it to devolve the inter-flake tunneling and intra-layer magnetotransport. Consequently, we determine the temperature dependences of the inter- and intra-layer characteristics, which both appear to be dominated by phonon-assisted processes at temperature $T > 20$ Kelvin. In particular, we identify the efficiency of phonon-assisted tunneling as the main limiting factor for electrical conductivity in graphene laminates at room temperature. We also demonstrate a thermoelectric sensitivity of around 50 μV K$^{-1}$ in a solution-processed metal-free graphene-based thermocouple.

The development of 2D material inks for printed electronics, supported by the expanding and perfected chemical exfoliation routes,[1] requires deeper insights into the factors limiting the electrical performance of the solution-processed 2D materials films[2]. The macroscopic models based on a phenomenological circuit theory[3, 4] imply that the overall resistance of a network of weakly coupled conductive nanosheets is defined by the interplay between their intrinsic conductivity and an interfacial 'contact' resistance.[5, 6] The mechanisms of electrical conductivity in the printed network of 2D nanoparticles have been recently probed by tracking the temperature dependence of resistivity.[7-11] For example, an exponential increase of resistivity upon cooling the film points towards thermally activated variable-range hopping transport between small semi-isolated grains.[9, 11-13] However, a variable-range hopping transport scenario hardly applies to a laminate composed of large-area metallic nanosheets with high intrinsic mobility.[10] Such materials have recently been studied, where the reported high values of optical conductivity increase upon cooling,[8, 10, 14] highlighted the role of phonons in electronic transport. At the same time, charge transport across the boundaries of nanosheets remains the main limiting factor for the film conductivity, especially in laminates produced from layered van der Waals materials,[15-20] which are sensitive to the implemented exfoliation chemistry and post-processing relative twist angle between the crystals, residually trapped intercalants, and packing density.[19, 21, 22]

Here, we study the graphene laminates produced by chlorosulphuric acid-assisted liquid phase exfoliation,[23] shown to deliver large aspect ratio (~$10^3$) high-quality nanosheets of graphene. In particular, we perform magnetotransport characterization of multiple solution-processed laminates with the same structural properties but different charge carrier types and doping densities controlled by the post-processing annealing. The analysis of magnetotransport data enables us to devolve the parametric dependences (temperature and carrier density) of the intra- and inter-flake transport. The application of a strong magnetic field leads to the distortion of the intra-nanosheet current flow towards their edges, forcing the carrier tunneling between the adjacent flakes to occur near the edges rather than across the whole overlap area, making the intralayer charge transport similar to that in a two-terminal Hall resistor.[24] In the reported experiment, this crossover was manifested as a transition of the observed magnetoresistance (*MR*) from a quadratic to the linear magnetic field dependence. The quantitative devolution of the measured parametric dependences was made using a mesoscale numerical model applied to a variety of networks composed of weakly coupled high-mobility 2D conductors, with the temperature dependences of the devolved transport parameters pointing towards the dominance of electron-phonon scattering processes in both intra-flake conductivity and inter-flake tunneling. Moreover, the performed XPS, Raman and X-ray diffraction characterization indicated that the only non-graphene chemical present in the laminates was the tetra-sulfonated pyrene (s-Py) molecules introduced during the exfoliation. Using the post-processing thermolysis for removing the sulfonic groups of the s-Py molecules, we modify both graphene doping and inter-flake tunnelling and optimize the material to achieve a substantial thermoelectric response, producing a highly sensitive all-graphene solution-processed thermocouple.

**Laminates of large aspect ratio bilayer graphene nanosheets**

All laminates discussed in this report were prepared by blade coating of a Polyethylene terephthalate (PET) film with an additive-free graphene slurry with a solid content of around 50 g·L$^{-1}$ in *N*-Methyl-2-pyrrolidone (NMP). The slurry was obtained from the dispersion of large-area graphene nanosheets in NMP produced by chlorosulfonic acid-assisted liquid phase exfoliation.[23] The nanosheets' lateral size covers a range between 1 μm and 20 μm with thicknesses ranging from 1 to around 40 graphene layers, Figures 1A and 1B. The final density of the dried and calendared laminates was between 1.6 and 2.0 g·cm$^{-3}$, depending on the quality and density of the precursor graphite crystals. The thickness of the final laminates shown in Figure 1C was in the range of 10 - 20 μm.

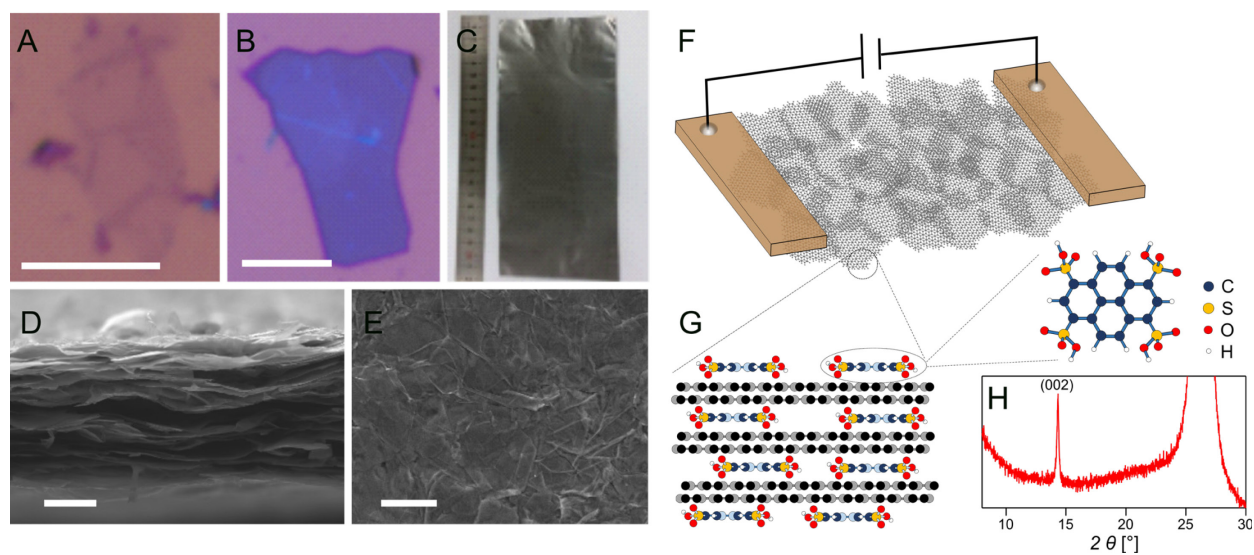

**Figure 1.** Graphene laminates made of large-area nanosheets. Panels A) and B) show a monolayer and 12.5 nm-thick graphene nanosheets deposited on a silicon wafer with the 290 nm oxide layer. The graphene nanosheets were assembled into a highly aligned laminate upon solution casting and mechanical compression C), where a high degree of alignment was observed in the SEM cross-section image, Panel D). Panel E) shows the top view of the laminates where randomly stacked nanosheets are visible. Panel (F) schematically shows the charge transport in such a randomly stacked network. Graphene nanosheets contain an organic molecule, tetra-sulfonated pyrene, mainly in the stage-2 intercalated region, Panel G).[23] The X-ray diffraction pattern (H) of the laminate confirms such a structure. Scale bars in A), B), D) and E) correspond to 5 μm.

The high aspect ratio of the nanosheets and the post-processing mechanical compression led to their alignment, which is essential for compact stacking and high flexibility.[5] The SEM cross-section images of the laminates, Figure 1D, confirmed the high degree of in-plane alignment of graphene nanosheets along with the complete twist misorientation evident from the top view SEM imaging, Figure 1E. Although the graphene slurry formulation is completely additive-free, the individual nanosheets are intercalated with tetra-sulfonated pyrene (s-Py), which is crucial for their large aspect ratio,[23] Figure 1G. The X-ray diffraction pattern obtained from these nanosheets, Figure 1H, appears to have the signature of the stage-II intercalation phase, suggesting that nanosheets comprise several aligned graphene bilayers, and the laminate is a collection of planarly stacked randomly twisted nanosheets with different overlap areas.

The doping type of the laminates was fine-tuned through controlled thermolysis of s-Py molecules through annealing in an inert atmosphere (Ar/H$_2$ or Ar) at up to 2200°C. Here, we exploited the electron-accepting nature of s-Py, which would be a natural *p*-type dopant for the as-prepared laminates[25] and the fact that the pristine pyrene is a weak electron donor (*n*-type dopant).[26] That makes it possible to adjust the doping type from the "as-prepared" *p*-type laminates to *n*-type laminates by controlled removal of the sulfonated groups through high-temperature annealing.

**Linear magnetoresistance in laminates of large aspect ratio bilayer graphene nanosheets**

The laminate samples used in magnetotransport experiments were shaped into millimeter-scale Hall bar structures by laser ablation and packaged inside standard ceramic chip carriers with the current driven in-plane and the magnetic field applied perpendicular to the laminate, Figure 2A. The measurements were performed between 1.6 K and room temperature in the presence of a magnetic field, $B$, up to 13.5 T.

The zero-field electrical conductivity, $\sigma_{xx}(B=0) = 1/\rho_{xx}$, for the as-prepared laminate (blue trace in Figure 2A) exhibited a weak insulator-like temperature dependence with a well-pronounced saturation region below ~20 K. The laminate annealed at 1100°C (red curve in Figure 2A) demonstrated a similar temperature dependence with a somewhat larger conductivity change and less pronounced saturation. Note that this temperature dependence is not strong enough to be described in terms of the hopping transport as was reported earlier.[9] We determined the room temperature carrier mobility using the measured low-field magnetoresistance, $MR =$ $[\rho_{xx}(B) - \rho_{xx}(B=0)]/\rho_{xx}(B=0)$, as $\mu = \sqrt{MR}/B \sim 1450$ cm$^2 \cdot$V$^{-1} \cdot$s$^{-1}$ for the as-prepared samples and ~2100 cm$^2 \cdot$V$^{-1} \cdot$s$^{-1}$ for annealed samples. At the same time, the experimentally found Hall conductivity, $\sigma_{xy} = -\rho_{xy}/[\rho_{xx}^2 + \rho_{xy}^2]$ in Figure 2B, indicates a change from *p*-type to *n*-type of roughly the sample doping concentration.

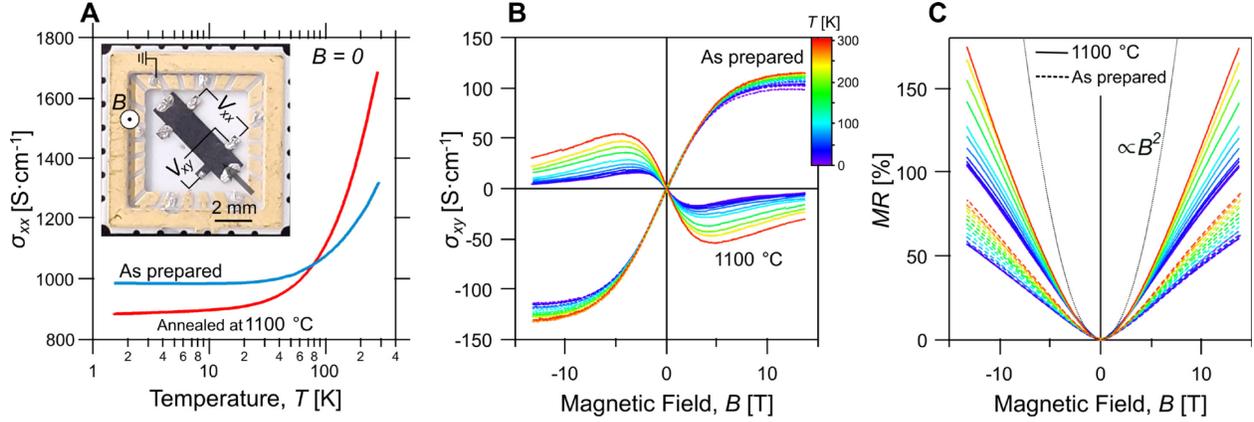

**Figure 2.** Transport characteristics of the as-prepared graphene and the laminate annealed at 1100°C. A) The temperature dependence of the zero-field conductivity $\sigma_{xx}$ of the as-prepared laminate (blue line), and after annealing at 1100°C (red line). Inset: the photograph of a typical laminate Hall bar device. B) The Hall conductivity $\sigma_{xy} = -\rho_{xy}/[\rho_{xx}^2 + \rho_{xy}^2]$ measured in the temperature range 1.6 K to 300 K. C) The magnetoresistance $MR = [\rho_{xx}(B) - \rho_{xx}(B=0)]/\rho_{xx}(B=0)$ for the as-prepared (dashed curves) and annealed at 1100°C (solid curves) laminates measured in the same temperature as in panel B). The quadratic response was visible in magnetoresistance at small magnetic fields ($MR \propto B^2$) as shown by the dotted line in panel C), and it was used to estimate the laminate's apparent carrier mobility as $\mu = \sqrt{MR}/B$ (see main text).

The striking feature of the measured $MR$, shown in Figure 2C, is the linear magnetoresistance characteristic for higher magnetic fields, $B > 3$ T, observed at all temperatures and in all samples. The linear $MR$, has been observed in several materials such as graphene,[27-31] silver chalcogenides,[32] indium antimonide,[33] Dirac semimetals,[34, 35] and topological semimetals.[36] In all of those by their own peculiar reasons. Here we point out that for all our samples with different post-processing annealing, the slope of the linear $MR$ decreases with the decreasing temperature, which we consider as a hint towards the relevance of the phonon-assisted processes in both in inta- and inter-layer transport characteristics.

Aiming at the quantitative analysis, we employ a model that accounts for the high intrinsic intra-layer carrier mobility and weak inter-layer coupling. For numerical simulations, we model a laminate as a network of overlapping nanosheets, each composed of $N_{BLG}$ aligned graphene bilayers (separated by intercalant molecules), and parametrized by the lateral conductivity tensor of individual nanosheets, $\widehat{\tilde{\sigma}}$, (with components $\tilde{\sigma}_{xx} = \tilde{\sigma}_{yy} = en\tilde{\mu}/(1 + (\tilde{\mu}B)^2)$ and $\tilde{\sigma}_{yx} = -\tilde{\sigma}_{xy} = en\tilde{\mu}^2 B/(1 + (\tilde{\mu}B)^2)$) and vertical tunnelling conductance per unit area of the overlapping regions, $s$, where $\tilde{\mu}$ - is the charge carrier mobility of a nanosheet, and $n = N_{BLG}\rho_{BLG}E_F$ is the carrier density per nanosheet composed of $N_{BLG} \approx 3.5$ bilayers with the density of states $\rho_{BLG}$ and the Fermi level $E_F$ counted from the bilayer's charge neutrality point.

While in the modeling we used various shapes of nanosheets, in Figure 3A, the typical modelling results are exemplified by an array of random overlapping polygons.

For a pair of overlapping nanosheets, denoted as "1" and "2", the intra- and inter-nanosheets current is related to potential distribution across the sheets as

$$\vec{\nabla} \cdot \vec{j}_1 = -\vec{\nabla} \cdot \vec{j}_2 = s(U_2(x,y) - U_1(x,y));$$
$$\vec{j}_{1,2} = -\widehat{\tilde{\sigma}}_{1,2} \vec{\nabla} \cdot U_{1,2}(x,y). \tag{1}$$

Here $U_{1,2}(x,y)$ and $\widehat{\tilde{\sigma}}_{1,2}$ -are the potential distribution and electrical conductivity of nanosheets 1 and 2 correspondingly. The two coupled equations in Equation (1) suggest a length scale, $l = \sqrt{\frac{\tilde{\sigma}_{xx}}{2s}}$, which characterizes the interlayer charge equilibration. We determine the overall resistivity of the film by solving numerically Eqs (1) for a periodic network, additionally averaging over a distribution of the Fermi energies of individual nanosheets with the mean value $\langle E_F \rangle$, and variance $\delta E_F$ (Note that in annealed samples, we have $\langle E_F \rangle \ll \delta E_F$). An example of current and potential distribution in several overlapping nanosheets at a high magnetic field such that $\tilde{\mu} B = 10$ is shown in Figure 3A.

Model in Eqs (1) allows us to develop an interpolation formula to describe the zero-field resistance of the laminate,

$$R = \frac{L}{2\tilde{\sigma}_{xx}} + \frac{l}{\tilde{\sigma}_{xx}} \coth\left(\frac{L}{2l}\right), \tag{2}$$

such that $R \approx \begin{cases} 1/(L\,s), & l \gg L \\ l/\tilde{\sigma}_{xx} = 1/\sqrt{2s\tilde{\sigma}_{xx}}, & l \ll L \end{cases}$ (see S.I. section S3 for details). This model also allows us to reproduce theoretically the linear magnetoresistance shown in Figure 2C. This appears to be the result of shrinking charge equilibration length upon increasing the magnetic field to the values where $\tilde{\mu}B \gg 1$, so that $l \sim \sqrt{\frac{en\tilde{\mu}}{2s(1+(\tilde{\mu}B)^2)}} \ll L$, which squeezes interlayer tunneling to hotspots with the size $\sim l$, Figures 3A and Figure S1. These considerations suggest that for large enough $B$, $R(B) \propto \sqrt{B/s}$ while at zero field $R(0) \propto 1/s$, leading to

$$MR = \frac{R(B)-R(B=0)}{R(B=0)} \propto \sqrt{s}B, \tag{3}$$

which slope is dependent on the interlayer coupling $s$ (for details, see Supporting Information, Figure S5). The linear behavior can also be argued as a network of two-terminal Hall devices.[37] The results of numerical $MR$ modeling are displayed in Figure 3B, where we compare a laminate built of random polygons with a periodic array of overlapping square-shaped nanosheets. In both realizations we find similar linear asymptotic $MR$ behavior (illustrated as $\partial MR/\partial B \to$ const), which reflects similar current density distribution and formation of the tunneling hot spots, Figure S1. Also, in Figure 3B we show two examples of the experimental data obtained from the high-temperature annealed laminates (for which $\langle E_F \rangle \ll \delta E_F$, and $s$ is used as the fitting parameter for each sample temperature). Based on this qualitative similarity, in the further data analysis we employ the square-shaped nanosheet model because of its higher numerical efficiency. While this simplified model gives only ~30% accuracy in the absolute values of the fitting parameters, it allows the establishment of their qualitative temperature dependences, as discussed in the next section.

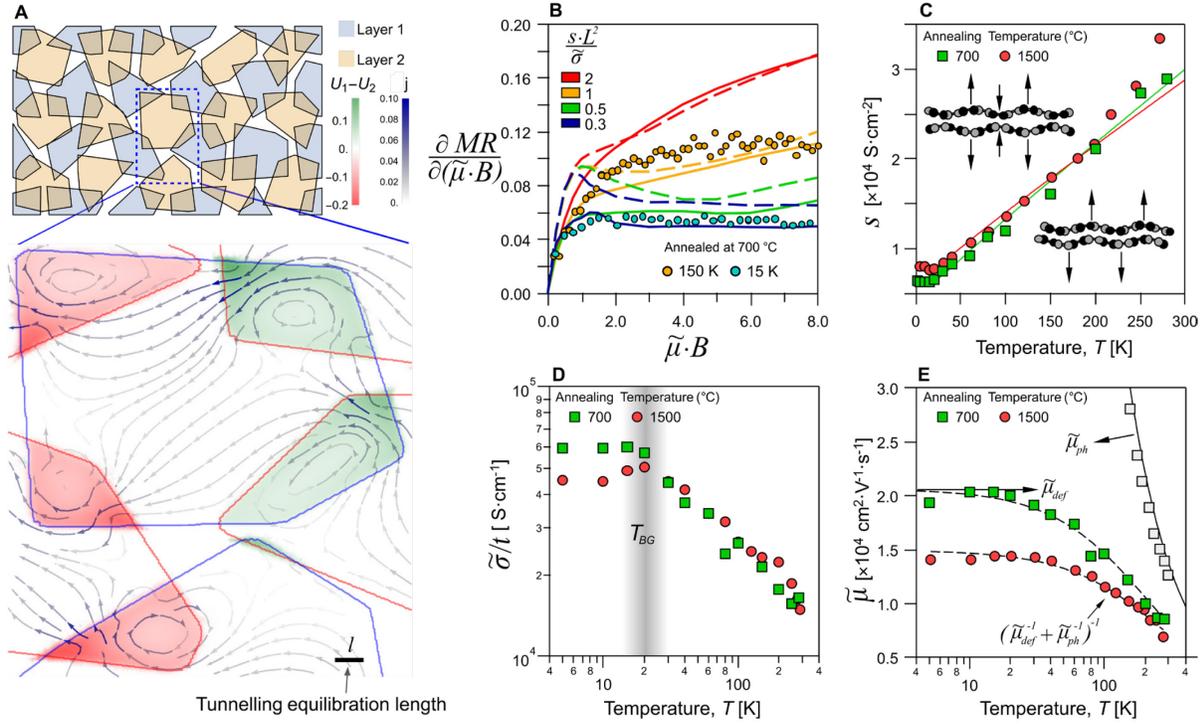

**Figure 3.** Numerical model of the linear magnetoresistance and transport parameters of individual graphene nanosheets. A) The results of the numerical modeling illustrating the spatial distribution of the potential difference $U_1 - U_2$ (color map) and total current (arrows) for two layers of polygons with the same $\tilde{\sigma}_{xx}$ conductivity but opposite $\tilde{\sigma}_{xy}$. The boundaries of the nanosheets are marked with blue and red curves. B) Comparison of MR derivative for the model of overlapping random polygons (solid) and periodic array of overlapping squares (dashed). Filled circles show the experimental data 15 K and 150 K from the annealed laminates extracted using the square lattice model. C) The temperature dependence of the tunneling interfacial conductance, $s$, (symbols) extracted from the experimental data measured for two samples annealed at 700°C (green symbols) and 1500°C (red symbols). Solid curves show the theoretical calculations made for the phonon-assisted tunneling model, with the coupling constant g being the only fitting parameter (see text). The insets in panel C) illustrate the graphene's beating

phonon modes considered for this calculation.[40] D) The individual nanosheets' conductivity, $\tilde{\sigma}$, where $t = 4$ nm for the same two laminates. The vertical band illustrates the Bloch-Gruneisen temperature, $T_{BG} \approx 20$ K. E) The individual nanosheets' carrier mobility for the laminates annealed at 700°C (green symbols) and 1500°C (red symbols). The dashed curves are the expected total carrier mobility defined as $\tilde{\mu}^{-1} = \tilde{\mu}_{def}^{-1} + \tilde{\mu}_{ph}^{-1}(T)$ with the same value of the temperature-dependent lattice contribution $\tilde{\mu}_{ph}(T)$. The grey symbols in E) show the mobility of graphite single crystals[38, 39] limited by the same lattice contribution as in our samples.

**Phonon-assisted tunneling and the electrical conductivity of individual nanosheets**

The results of fitting of the experimentally observed $MR$ – from weak magnetic fields where it is quadratic to the high-field linear regime – with $\tilde{\mu}$ and $s$ used as fitting parameters at each temperature (*e.g.* at 5 K and 300 K; see Figure S3 for specific examples) are shown in Figure 3C-E. This gives us information about the temperature dependencies of the interlayer tunneling, Figure 3C, individual nanosheet conductivity, Figure 3D, and the intrinsic mobility of the constituent bilayers, Figure 3E.

The temperature dependence of the nanosheet mobility $\tilde{\mu}$ (symbols in Figure 3E) is typical for a charge transport limited by defects at low temperatures and electron-phonon scattering at higher temperatures, $\tilde{\mu}^{-1} = \tilde{\mu}_{def}^{-1} + \tilde{\mu}_{ph}^{-1}(T)$, where $\tilde{\mu}_{def}$ – is the defect-limited mobility found from the low-temperature saturation region and $\tilde{\mu}_{ph}(T)$ – is the temperature-dependent phonon-limited contribution. For all samples used in this study, the phonon-limited component appears to be universal and described as $\tilde{\mu}_{ph}(T) = 1275/T^{1.2}$ [m²·V⁻¹·s⁻¹] (solid line in Figure 3E) – in a reasonable agreement with the values previously reported[38, 39] for bulk graphite (open squares). This phonon-limited room-temperature intrinsic carrier mobility is a sign of a high crystallinity of the constituent nanosheets and remarkable micro-scale homogeneity of the s-Py intercalant distribution.

Moreover, in all tested samples, the extracted value of the interlayer tunneling conductivity, $s$, acquires distinct linear asymptotic upon the temperature increase above 20 K, Figure 3C. We attribute this behavior to the fact that the tunneling between misaligned graphene nanosheets requires a substantial momentum transfer due to the mismatch between $K$-points in the Brillouin zones of consecutive randomly oriented bilayers. While at low temperatures this interlayer momentum mismatch can only be overcome by disorder-assisted tunneling (*e.g.* resonant tunneling through the s-Py orbitals), at elevated temperatures, the inelastic phonon-assisted processes kick in, compensating the momentum mismatch by the momentum transfer from or to graphene's breathing phonon modes. Tunneling assisted by the breathing phonon modes has been previously identified as the primary transport mechanism between strongly twisted monolayer graphene flakes.[40] To describe the contribution of the out-of-plane vibrations of graphene, we use the phonon-assisted tunneling conductance derived in Supporting Information, section S4

$$s(\theta) = \frac{e^2 g^2 m_{\text{BLG}}^2}{4\pi\hbar^3 \rho_C k_B T} \left[\sinh\left(\frac{\hbar\omega_{q_K}}{2 k_B T}\right)\right]^{-2}, \tag{4}$$

which takes into account both spontaneous and stimulated emission/absorption of phonons. Here $\rho_C = 7.6\times10^{-7}$ kg·m$^{-2}$ is the mass density of graphene, $m_{\text{BLG}} \approx 0.03\, m_e$ is the effective mass for electrons in BLG used to estimate its density of states and $\omega_{q_K}$ is the energy of the beating phonon mode at momentum $q = 2K\sin(\theta/2)$, connecting the K-points ($K = 4\frac{\pi}{3a} \approx 17$ nm$^{-1}$) of graphenes twisted by angle $\theta$. As shown in Figure S6, the randomness of the twist angle in the network can be accounted for by a simplified model with the same tunneling between all sheets with the twist angle fixed to $\theta \approx 16°$. The onset of a linear-like dependence of $s$ versus $T$ occurs at $T \approx 0.2\, \hbar\omega_q/k_B$, which suggests that both the breather and flexural modes of BLG are engaged in the tunneling conductivity and have energies at around $\hbar\omega_q \approx 10$ meV. While the temperature dependence of $T$ resembles the one observed in twisted bilayers of graphene,[40] its lower absolute value in Figure 3C can be attributed to a larger separation between the bilayers caused by the presence of s-Py molecules.

**All-graphene bipolar thermocouple**

As noted above, the post-processing annealing at 1100°C leads to the reversal of doping from p-type to n-type, evidenced by the Hall conductivity sign change (Figure 2B). To gain more information about the doping mechanism, we correlated the thermoelectric measurements with the X-ray photoelectron spectroscopy (XPS) analysis and Raman spectroscopy. Figure 4A combines the data for thermoelectric power and the XPS oxygen-to-carbon content ratio (O/C) found in laminates annealed at different temperatures. The thermopower correlates well with the O/C ratio up to around 1100°C. Initially, both remain unchanged to $T_1 \approx 350°C$ and then start to decrease – the thermopower from +42 µV·K$^{-1}$ to about –14 µV·K$^{-1}$ and the O/C ratio from 6.7% to 1.6% due to the removal of sulfur-bounded oxygen (Figure S9). The onset at $T_1$ is attributed to the carbon-sulfur bond cleavage reaction,[41] and signifies the desulfonation of the s-Py acceptor molecules[25] through the thermolysis reaction. The desulfonation of s-Py causes a gradual change from the p-type conductivity to n-type when the pyrene is stripped off from the sulfonic groups

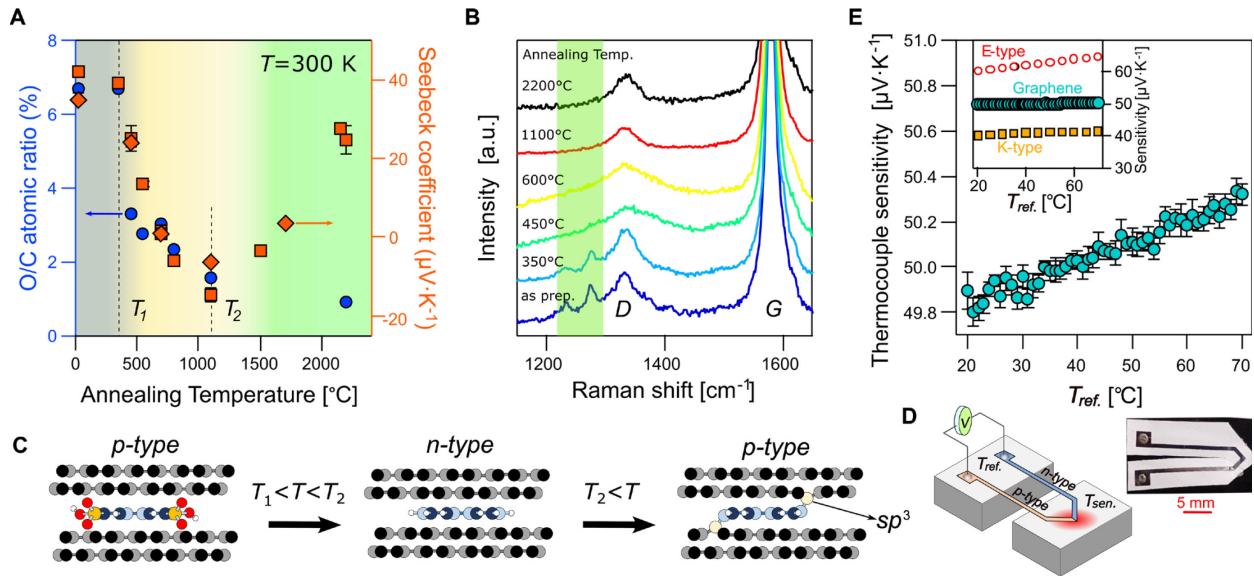

**Figure 4.** Tuning the thermopower of graphene laminates by s-Py charge transfer. A) The correlation between the Seebeck coefficient (orange symbols) and the oxygen-to-carbon atomic ratio obtained from the XPS spectrum (blue circles) for the laminates annealed at different temperatures. Temperatures $T_1 \approx 350°C$ and $T_2 \approx 1100°C$ mark the onset of the desulfonation and dehydrogenation of an s-Py molecule. Orange diamond and square symbols denote the thermopower data obtained from the laminates derived from different graphite materials (Ashbury Graphite and Graphenium flakes, NGS). B) The Raman spectra of graphene laminates annealed at different temperatures demonstrate the disappearance of the band between 1230 cm$^{-1}$ and 1275 cm$^{-1}$ (highlighted in green) associated with the s-Py molecules. C) The schematics illustrate the thermolysis of s-Py, which is responsible for the reversal of the conductivity from p-type to n-type at low annealing temperatures ($T_1 < T < T_2$). At higher annealing temperatures ($T < T_2$), the dehydrogenation and formation of the out-of-plane sp$^3$ carbon bonds responsible for the reentrant p-type conductivity. D) The schematics of the all-graphene thermocouple device are made of two graphene laminate strips with the individual Seebeck coefficients tuned by annealing. Inset: The photograph of the actual thermocouple device. E) Thermocouple sensitivity measured at different

reference temperatures, $T_{\text{ref}}$ as illustrated in panel D). The sensitivity is comparable to the commercial K-type and E-type thermocouples, as shown in the inset. (NIST database, https://srdata.nist.gov/its90/main/).

and acts as a donor dopant.[26] The loss of the sulfonic groups is also confirmed by the Raman spectroscopy (Figure 4B), showing a gradual disappearance of the bands between 1200 cm$^{-1}$ and 1300 cm$^{-1}$ associated with s-Py.[42] For the samples annealed at $T_2 \approx 1100°C$ and higher, the sulfur content was undetectable in XPS signal (Figure S9).

Further annealing above $T_2 \approx 1100°C$ returns *p*-type doping of laminates, despite the saturated O/C ratio (Figure 4A, Figure S10). We suggest that at such high temperatures, pyrene molecules undergo cleavage of their carbon-hydrogen bonds and formation of new $\sigma$- bonds with the carbon atoms from the adjacent graphene layer, inflicting some *sp*$^3$-hybridization,[43] Figure 4C. The electron deficiency created in these new *sp*$^3$ bonds leads to overall *p*-type doping of graphene, which explains the observed reversal of the Seebeck coefficient in material annealed at higher temperatures, Figure 4C. At the same time, the formation of the out-of-plane *sp*$^3$ bonds produces point-like defects, which show up in a ~30% increase in the D peak intensity in Raman and a proportional drop in the carrier mobility, Figure 3E.

All these give access to the controlled tuning of laminates' Seebeck coefficient by the post-processing annealing. The thermopower values and ability to vary its sign we achieve in laminates, *e.g.*, as prepared and annealed at 1100°C, are comparable to those of commonly used metal thermocouple alloys. To demonstrate the potential application of such laminates, we fabricated a proof-of-principle all-graphene thermocouple made of combined *p*- and *n*-type laminates, Figure 4D, as-prepared and annealed at 1100°C, respectively. The fabricated thermocouple performs with a sensitivity of ~50 µV·K$^{-1}$, Figure 4E. With reference to the second reversal of the doping type at higher annealing temperatures, the result in Figure 4A suggests that one can make an equally sensitive thermocouple using the 1100°C and 2200°C annealed laminates which would be safe to operate up to 1000°C.

**Conclusion**

This work showed that the graphene laminates prepared by the acid-assisted liquid-phase exfoliation technique followed by a high-temperature annealing result in high electrical conductivity characterised by a linear $MR$ regime in a strong magnetic field ($\mu B \gg 1$). The linear $MR$ regime is explained by a combination of tunneling coupling between the nanosheets and strong local Hall currents, both of which contribute to the total conductivity of the nanosheet network. Equipped with the $MR$ data and numerical modeling, we disentangle the contribution of the intra-plane and inter-plane conductivities. The observed temperature dependences of both quantities point to the electron-phonon coupling dominated transport at a temperature above 100 K, displaying the high intrinsic quality of the constituent nanosheets. We also demonstrated the unique ability of the described laminates to change their doping, both value and polarity, depending on the post-processing annealing temperature, which we used to produce a proof-of-principle all-graphene thermocouple with competitive sensitivity. Overall, the report opens new avenues towards solution-processable materials for sensors and devices.

**Experimental Methods**

Dispersions of graphene in *N*-Methyl-2-pyrrolidone (NMP) were obtained through a chlorosulfonic acid-assisted exfoliation method, as previously described in our work.[23] Our analysis revealed that the median size and thickness of the nanosheets were 5.6 μm and 4.0 nm, respectively.[25] The dispersions were centrifugated at 12,000 rpm for one hour to separate the nanosheets from the acid and excess pyrene. The resulting sediments were collected and redispersed in NMP to form a homogeneous slurry containing 5 wt.% of graphene.

This viscous slurry was then directly cast onto a plastic (PET) substrate using the doctor blade coating technique without adding any other substances. The laminates were dried at 90°C for two days. Since NMP has a high boiling temperature, the remaining solvent was removed by immersing the dried film in a hot water bath. The water removed most of the residual NMP, and the film was subsequently dried at 90°C overnight. To further increase the film's density, the samples were mechanically compressed using a twin roller press and then manually peeled off from the substrate.

Subsequently, the samples were annealed in an electric furnace (Carbolite quartz tube furnace) under an Ar/$H_2$ atmosphere at the target temperature for 2 hours, with a heating and cooling rate of 2°C·$min^{-1}$. For samples annealed at temperatures above 1100°C, a different furnace (FCT Systeme GmbH) was used in an Ar atmosphere, maintaining the same annealing rate and duration. After annealing, all samples underwent calendering in a hydraulic press at 350 bar.

Scanning electron microscopy (Tescan Mira 3 FEG-SEM) or optical microscopy (Nikon Eclipse LV100ND) was employed to determine the samples' thickness.

The elemental composition of samples was evaluated using XPS on an ESCA2SR spectrometer (Scienta Omicron GmbH) or HAPEX (HAXPES-Lab, Scienta Omicron GmbH). Raman spectra were captured on Renishaw inVia using a 633 nm laser. X-ray diffraction patterns were collected on a Rigaku SmartLab diffractometer using Cu K-α radiation.

The samples were cut into Hall bar geometry, 1.5 mm × 3.0 mm (between the potential probes), using the laser photo-ablation system (M-Solv MSV-100) and then bonded inside a ceramic chip carrier (see inset of Figure 2A). The transport measurements were performed in a cryogen-free variable temperature cryostat equipped with a 14 Tesla superconducting magnet (Cryogenic Limited, CFMS-14T). To determine the longitudinal, $\rho_{xx}$, and transversal, $\rho_{xy}$, resistivities, the longitudinal, $V_{xx}$, and Hall, $V_{xy}$ voltages were measured by applying a current of $I = 10$ µA using a lock-in amplifier (Stanford Research Systems SR860). The bulk conductivity tensor was then calculated from the geometry and thickness of the samples.

The Seebeck coefficient of the samples was measured on a home-built setup at room temperature. In brief, the samples were mounted between two aluminum posts with individual PID temperature control achieved with miniature Peltier modules (European Thermodynamics, APH-031-10-13-S) and a pair of high-precision platinum RTD (Innovative Sensor Technology, P0K1.232.4W.K.010). The dual-channel source meter units (Keithley 2636B) operated with LabView software provided the temperature reading and heating power supply, and the low-noise nanovoltmeter (Keithley 2182A) was used to measure the thermoelectric voltage.

ASSOCIATED CONTENT

**Supporting Information**.

The Supporting Information is available free of charge and contains a detailed description of the conductivity numerical model, a discussion of the nanosheet electrical properties obtained using this model, an expanded discussion of the origins of the linear magnetotransport in a graphene laminate, a detailed theoretical description of the phonon-assisted tunnelling processes, an assessment of the number of the point defects in individual nanosheets using the Raman spectrometry and the XPS compositional analysis of the laminates used in this study.

AUTHOR INFORMATION

Corresponding Author

*Andrey V. Kretinin. E-mail: andrey.kretinin@manchester.ac.uk

Author Contributions

The manuscript was written with contributions from all authors. All authors have approved the final version of the manuscript.

Funding Sources

This work was supported by EC FET Core 3 European Graphene Flagship Project, EPSRC grants number EP/S030719/1, EP/V007033/1, EP/V036343/1 and the Lloyd Register Foundation Nanotechnology Grant.

ACKNOWLEDGMENT

We are grateful to K. S. Novoselov and A. K. Geim for their encouragement at the early stages of this work. We also thank H. Woods, A. Brook, P. Higgins, and P. Greensmith for their technical assistance.

REFERENCES

1. Pinilla, S.; Coelho, J.; Li, K.; Liu, J.; Nicolosi, V., Two-dimensional material inks. *Nature Reviews Materials* **2022,** *7* (9), 717-735.

2. Lin, Z.; Liu, Y.; Halim, U.; Ding, M.; Liu, Y.; Wang, Y.; Jia, C.; Chen, P.; Duan, X.; Wang, C.; Song, F.; Li, M.; Wan, C.; Huang, Y.; Duan, X., Solution-processable 2D semiconductors for high-performance large-area electronics. *Nature* **2018,** *562* (7726), 254-258.

3. Yao, H.; Hsieh, Y.-P.; Kong, J.; Hofmann, M., Modelling electrical conduction in nanostructure assemblies through complex networks. *Nature Materials* **2020,** *19* (7), 745-751.

4. Wang, F.; Gosling, J. H.; Trindade, G. F.; Rance, G. A.; Makarovsky, O.; Cottam, N. D.; Kudrynskyi, Z.; Balanov, A. G.; Greenaway, M. T.; Wildman, R. D., Inter-flake quantum transport of electrons and holes in inkjet-printed graphene devices. *Advanced Functional Materials* **2021,** *31* (5), 2007478.

5. Kelly, A. G.; O'Suilleabhain, D.; Gabbett, C.; Coleman, J. N., The electrical conductivity of solution-processed nanosheet networks. *Nature Reviews Materials* **2022,** *7* (3), 217-234.

6. Clifford, K.; Ogilvie, S. P.; Amorim Graf, A.; Wood, H. J.; Sehnal, A. C.; Salvage, J. P.; Lynch, P. J.; Large, M. J.; Dalton, A. B., Emergent high conductivity in size-selected graphene networks. *Carbon* **2024,** *218*, 118642.

7. Ippolito, S.; Kelly, A. G.; Furlan de Oliveira, R.; Stoeckel, M.-A.; Iglesias, D.; Roy, A.; Downing, C.; Bian, Z.; Lombardi, L.; Samad, Y. A.; Nicolosi, V.; Ferrari, A. C.; Coleman, J. N.; Samorì, P., Covalently interconnected transition metal dichalcogenide networks via defect engineering for high-performance electronic devices. *Nature Nanotechnology* **2021,** *16* (5), 592-598.

8. Liu, Z.; Zhang, H.; Eredia, M.; Qiu, H.; Baaziz, W.; Ersen, O.; Ciesielski, A.; Bonn, M.; Wang, H. I.; Samorì, P., Water-Dispersed High-Quality Graphene: A Green Solution for Efficient Energy Storage Applications. *ACS Nano* **2019,** *13* (8), 9431-9441.

9. Piatti, E.; Arbab, A.; Galanti, F.; Carey, T.; Anzi, L.; Spurling, D.; Roy, A.; Zhussupbekova, A.; Patel, K. A.; Kim, J. M.; Daghero, D.; Sordan, R.; Nicolosi, V.; Gonnelli, R. S.; Torrisi, F., Charge transport mechanisms in inkjet-printed thin-film transistors based on two-dimensional materials. *Nature Electronics* **2021,** *4* (12), 893-905.

10. Zheng, W.; Sun, B.; Li, D.; Gali, S. M.; Zhang, H.; Fu, S.; Di Virgilio, L.; Li, Z.; Yang, S.; Zhou, S.; Beljonne, D.; Yu, M.; Feng, X.; Wang, H. I.; Bonn, M., Band transport by large Fröhlich polarons in MXenes. *Nature Physics* **2022,** *18* (5), 544-550.

11. Cottam, N. D.; Wang, F.; Austin, J. S.; Tuck, C. J.; Hague, R.; Fromhold, M.; Escoffier, W.; Goiran, M.; Pierre, M.; Makarovsky, O.; Turyanska, L., Quantum Nature of Charge Transport in Inkjet-Printed Graphene Revealed in High Magnetic Fields up to 60T. *Small* **2024,** 2311416.

12. Efros, A. L.; Shklovskii, B. I., Coulomb gap and low temperature conductivity of disordered systems. *Journal of Physics C: Solid State Physics* **1975,** *8* (4), L49.

13. Park, M.; Hong, S. J.; Kim, K. H.; Kang, H.; Lee, M.; Jeong, D. H.; Park, Y. W.; Kim, B. H., Electrical and thermoelectric transport by variable range hopping in reduced graphene oxide. *Applied Physics Letters* **2017,** *111* (17), 173103.

14. Dong, R.; Han, P.; Arora, H.; Ballabio, M.; Karakus, M.; Zhang, Z.; Shekhar, C.; Adler, P.; Petkov, P. S.; Erbe, A.; Mannsfeld, S. C. B.; Felser, C.; Heine, T.; Bonn, M.; Feng, X.; Cánovas, E., High-mobility band-like charge transport in a semiconducting two-dimensional metal–organic framework. *Nature Materials* **2018,** *17* (11), 1027-1032.

15. Allain, A.; Kang, J.; Banerjee, K.; Kis, A., Electrical contacts to two-dimensional semiconductors. *Nature Materials* **2015,** *14* (12), 1195-1205.

16. Cummings, A. W.; Duong, D. L.; Nguyen, V. L.; Van Tuan, D.; Kotakoski, J.; Barrios Vargas, J. E.; Lee, Y. H.; Roche, S., Charge transport in polycrystalline graphene: challenges and opportunities. *Advanced Materials* **2014,** *26* (30), 5079-5094.

17. Kim, T.-H.; Zhang, X.-G.; Nicholson, D. M.; Evans, B. M.; Kulkarni, N. S.; Radhakrishnan, B.; Kenik, E. A.; Li, A.-P., Large discrete resistance jump at grain boundary in copper nanowire. *Nano letters* **2010,** *10* (8), 3096-3100.

18. Kim, Y.; Yun, H.; Nam, S.-G.; Son, M.; Lee, D. S.; Kim, D. C.; Seo, S.; Choi, H. C.; Lee, H.-J.; Lee, S. W., Breakdown of the interlayer coherence in twisted bilayer graphene. *Physical Review Letters* **2013,** *110* (9), 096602.

19. Koren, E.; Leven, I.; Lörtscher, E.; Knoll, A.; Hod, O.; Duerig, U., Coherent commensurate electronic states at the interface between misoriented graphene layers. *Nature Nanotechnology* **2016,** *11* (9), 752-757.

20. Inbar, A.; Birkbeck, J.; Xiao, J.; Taniguchi, T.; Watanabe, K.; Yan, B.; Oreg, Y.; Stern, A.; Berg, E.; Ilani, S., The quantum twisting microscope. *Nature* **2023,** *614* (7949), 682-687.

21. Maslov, D. L.; Yudson, V. I.; Somoza, A. M.; Ortuño, M., Delocalization by Disorder in Layered Systems. *Physical Review Letters* **2009,** *102* (21), 216601.

22. Lin, Z.; Huang, Y.; Duan, X., Van der Waals thin-film electronics. *Nature Electronics* **2019,** *2* (9), 378-388.

23. Moazzami Gudarzi, M.; Asaad, M.; Mao, B.; Pinter, G.; Guo, J.; Smith, M.; Zhong, X.; Georgiou, T.; Gorbachev, R.; Haigh, S. J.; Novoselov, K. S.; Kretinin, A. V., Chlorosulfuric acid-assisted production of functional 2D materials. *npj 2D Materials and Applications* **2021,** *5* (1), 35.

24. Alekseev, P.; Dmitriev, A.; Gornyi, I.; Kachorovskii, V.; Narozhny, B.; Schütt, M.; Titov, M., Magnetoresistance in Two-Component Systems. *Physical Review Letters* **2015,** *114* (15).

25. Dong, X.; Fu, D.; Fang, W.; Shi, Y.; Chen, P.; Li, L. J., Doping single-layer graphene with aromatic molecules. *Small* **2009,** *5* (12), 1422-1426.

26. Kawano, S.-i.; Baumgarten, M.; Chercka, D.; Enkelmann, V.; Müllen, K., Electron donors and acceptors based on 2, 7-functionalized pyrene-4, 5, 9, 10-tetraone. *Chemical Communications* **2013,** *49* (44), 5058-5060.

27. Dykhne, A., Anomalous Plasma Resistance in a Strong Magnetic Field. *Soviet Physics Jetp-Ussr* **1971,** *32* (2), 348.

28. Ping, J.; Yudhistira, I.; Ramakrishnan, N.; Cho, S.; Adam, S.; Fuhrer, M. S., Disorder-Induced Magnetoresistance in a Two-Dimensional Electron System. *Physical Review Letters* **2014,** *113* (4), 047206.

29. Kisslinger, F.; Ott, C.; Heide, C.; Kampert, E.; Butz, B.; Spiecker, E.; Shallcross, S.; Weber, H. B., Linear magnetoresistance in mosaic-like bilayer graphene. *Nature Physics* **2015,** *11* (8), 650-653.

30. Vasileva, G. Y.; Smirnov, D.; Ivanov, Y. L.; Vasilyev, Y. B.; Alekseev, P.; Dmitriev, A.; Gornyi, I.; Kachorovskii, V. Y.; Titov, M.; Narozhny, B., Linear magnetoresistance in compensated graphene bilayer. *Physical Review B* **2016,** *93* (19), 195430.

31. Xin, N.; Lourembam, J.; Kumaravadivel, P.; Kazantsev, A. E.; Wu, Z.; Mullan, C.; Barrier, J.; Geim, A. A.; Grigorieva, I. V.; Mishchenko, A.; Principi, A.; Fal'ko, V. I.; Ponomarenko, L. A.; Geim, A. K.; Berdyugin, A. I., Giant magnetoresistance of Dirac plasma in high-mobility graphene. *Nature* **2023,** *616* (7956), 270-274.

32. Xu, R.; Husmann, A.; Rosenbaum, T. F.; Saboungi, M. L.; Enderby, J. E.; Littlewood, P. B., Large magnetoresistance in non-magnetic silver chalcogenides. *Nature* **1997,** *390* (6655), 57-60.

33. Hu, J.; Rosenbaum, T. F., Classical and quantum routes to linear magnetoresistance. *Nature Materials* **2008,** *7* (9), 697-700.

34. Singha, R.; Pariari, A. K.; Satpati, B.; Mandal, P., Large nonsaturating magnetoresistance and signature of nondegenerate Dirac nodes in ZrSiS. *Proceedings of the National Academy of Sciences* **2017,** *114* (10), 2468-2473.

35. Narayanan, A.; Watson, M.; Blake, S.; Bruyant, N.; Drigo, L.; Chen, Y.; Prabhakaran, D.; Yan, B.; Felser, C.; Kong, T., Linear magnetoresistance caused by mobility fluctuations in n-doped $Cd_3As_2$. *Physical Review Letters* **2015,** *114* (11), 117201.

36. Wang, A.-Q.; Ye, X.-G.; Yu, D.-P.; Liao, Z.-M., Topological Semimetal Nanostructures: From Properties to Topotronics. *ACS Nano* **2020,** *14* (4), 3755-3778.

37. Parish, M.; Littlewood, P., Non-saturating magnetoresistance in heavily disordered semiconductors. *Nature* **2003,** *426* (6963), 162-165.

38. Sugihara, K.; Kawamuma, K.; Tsuzuku, T., Temperature dependence of the average mobility in graphite. *Journal of the Physical Society of Japan* **1979,** *47* (4), 1210-1215.

39. Soule, D., Magnetic field dependence of the Hall effect and magnetoresistance in graphite single crystals. *Physical Review* **1958,** *112* (3), 698.

40. Perebeinos, V.; Tersoff, J.; Avouris, P., Phonon-Mediated Interlayer Conductance in Twisted Graphene Bilayers. *Physical Review Letters* **2012,** *109* (23), 236604.

41. Jiang, D. D.; Yao, Q.; McKinney, M. A.; Wilkie, C. A., TGA/FTIR studies on the thermal degradation of some polymeric sulfonic and phosphonic acids and their sodium salts. *Polymer degradation and stability* **1999,** *63* (3), 423-434.

42. Yang, H.; Schlierf, A.; Felten, A.; Eckmann, A.; Johal, S.; Louette, P.; Pireaux, J.-J.; Mullen, K., A simple method for graphene production based on exfoliation of graphite in water using 1-pyrenesulfonic acid sodium salt. *Carbon* **2013,** *53*, 357-365.

43. Telling, R. H.; Ewels, C. P.; El-Barbary, A. A.; Heggie, M. I., Wigner defects bridge the graphite gap. *Nature Materials* **2003,** *2* (5), 333-337.

Supporting Information# Ultimate charge transport regimes in doping-controlled graphene laminates: phonon-assisted processes revealed by the linear magnetoresistance.

*Mohsen Moazzami Gudarzi [1], Sergey Slizovskiy [1], Boyang Mao [1,3], Endre Tovari [4], Gergo Pinter [2], David Sanderson [2], Maryana Asad [2], Ying Xiang [2], Zhiyuan Wang [2], Jianqiang Guo [2], Ben F. Spencer [2], Alexandra A. Geim [5], Vladimir I. Fal'ko [1,6,7] and Andrey V. Kretinin*[,1,2,6]*

* E-mail: andrey.kretinin@manchester.ac.uk

[1] Department of Physics and Astronomy, The University of Manchester, Oxford Road, Manchester, M13 9PL, UK

[2] Department of Materials, The University of Manchester, Oxford Road, Manchester, M13 9PL, UK

[3] Cambridge Graphene Centre, Department of Engineering, University of Cambridge, 9 JJ Thomson Ave, Cambridge CB3 0FA

[4] Department of Physics, Institute of Physics, Budapest University of Technology and Economics, Műegyetem rkp. 3, H-1111 Budapest, Hungary
27

[5] Department of Physics, Harvard University, Cambridge, MA, USA

[6] National Graphgene Institute, The University of Manchester, Oxford Road, Manchester, M13 9PL, UK

[7] Henry Royce Institute for Advanced Materials, The University of Manchester, Oxford Road, Manchester, M13 9PL, UK

### S1. Numerical model for conductivity in laminate

As discussed in the main text and illustrated in Figure 1G, the laminate is made of few-layer graphene nanosheets, which are intercalated with pyrene molecules at every second layer (stage II intercalation), effectively splitting the multilayer flakes into a collection of aligned bilayers. The electrical conductivity between the aligned bilayers is good since they are aligned, while the conductivity between different misaligned flakes is incoherent and is governed by impurity-assisted and phonon-assisted tunneling mechanisms. Experimental data indicate a weak dependence of resistivity on the annealing temperature when the Hall resistance changes its sign. Such behavior indicates an ambipolar conduction mechanism involving carriers with a different sign of electric charge. However, the bilayer graphene has only one carrier type at a specified Fermi level: either electrons or holes. Coupling between the bilayers can induce ambipolar conduction, but this effect would be absent in the thinner flakes and would have to originate from the hopping process to the next nearest layer, which has a typical energy of $\gamma_2 \approx 20$ meV in graphite, and we expect a much smaller value in intercalated flakes due to the larger vertical separation between bilayers. An alternative approach is to assume that all the flakes have an electronic dispersion of several weakly coupled graphene bilayers, but different flakes have different Fermi levels (relative to the charge-neutrality point). To further simplify the model, we assume that each flake consists of $N_{\text{BLG}} \approx 3.5$ bilayers (corresponding to the average flake thickness found in the experiment) share the same Fermi level. The Fermi level positions are assumed to be normally distributed across the set of flakes with dispersion $\delta E_{\text{F}}$ and expectation value $\langle E_{\text{F}} \rangle$. The flakes are chosen to be of a square $L \times L$ ($L = 0.8$ in the numerical model) shape and tile two layers with $0.2L$ gaps, see Figure S1A. In the regions where flakes overlap, there is a vertical tunneling conductivity per unit area $s$.

We have also verified that the model utilizing random polygons, Figure S1B, produces qualitatively similar results despite notably smaller overlap areas.

The linear magnetoresistance ($MR$) in our model appears from overlapping flakes being connected in a two-terminal configuration, and the $MR$ is only weakly sensitive to the relative sign of the Hall coefficient of overlapping flakes. In Figure S1, we assumed an opposite (*p*- and *n*-) doping of the overlapping flakes, and in Figure S2 we compare the cases of the opposite and the same doping, showing that both arrangements result in almost the same $MR$ response.

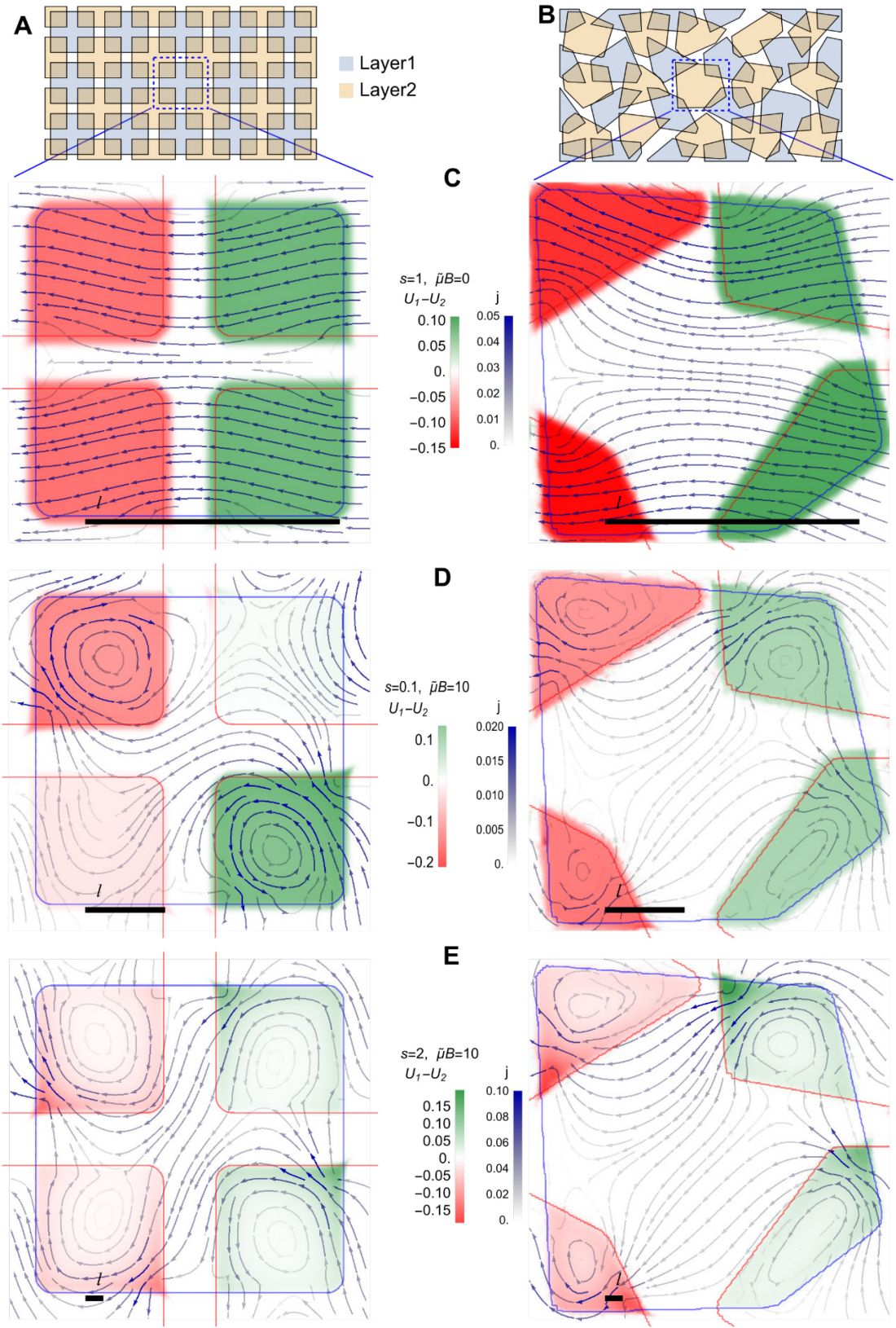

**Figure S1**. Panels A and B: Nanosheet network geometry of flakes used in numerical modeling. Results for two layers of oppositely-doped squares (left) and polygons (right) are shown. Panels C, D and E: examples of total current distribution (arrows) and the potential difference between the layers (red-to-green color scale) for $B = 0$ (C) and $\tilde{\mu}B = 10$ (D and E). The black scale bar corresponds to tunneling equilibration length $l = \sqrt{\frac{en\tilde{\mu}}{2s(1+(\tilde{\mu}B)^2)}}$, which is comparable to flake size in (C, D) and notably shorter in (E). Dimensionless tunneling $s$ corresponds to $s_{\text{model}}$ as given in the text. Overlapping flakes are assumed to have opposite (*p*- and *n*-) doping.

The flake carrier mobilities are chosen as $\tilde{\mu}$ for electrons and $0.9\,\tilde{\mu}$ for holes (when $E_F < 0$) to account for a larger effective mass of holes in bilayer graphene and graphite.[1] We use the Drude formula for in-plane flake conductivity $\tilde{\sigma}_{xx} + i\,\tilde{\sigma}_{xy} = \frac{\sigma_0}{1+i\tilde{\mu}B}$ where $\sigma_0 = N_{\text{BLG}}\rho_{\text{BLG}} e\,\tilde{\mu}\,E_F$ is determined by the Fermi level that is randomly chosen for each flake from the normal distribution. The physical vertical conductivity can be represented in the form of a dimensionless parameter

$$s_{\text{model}} = \frac{s\,L^2}{\tilde{\sigma}_{\text{typical}}}, \text{where } \tilde{\sigma}_{\text{typical}} = N_{\text{BLG}}\rho_{\text{BLG}} e\,\tilde{\mu}\,\delta E_F$$

is the value of in-plane conductivity determined by the typical width, $\delta E_F$, of Fermi level distribution.

For the two-layer model system we solve the current conservation equations numerically to find a potential distribution, $U_{1,2}$, where the current flux between the flakes is proportional to the voltage difference:

$$\vec{\nabla} \cdot \vec{j}_1 = -\vec{\nabla} \cdot \vec{j}_2 = s_{1-2}\big(U_2(x,y) - U_1(x,y)\big)$$
$$\vec{j}_{1,2} = -\tilde{\sigma}_{1,2}\vec{\nabla} \cdot U_{1,2}(x,y) \tag{S1}$$

magnetoresistance $MR = \frac{\rho_{xx}(B) - \rho_{xx}(0)}{\rho_{xx}(0)}$ and normalized Hall conductivity $\frac{\sigma_{xy}(B)}{\sigma_{xx}(B=0)}$ are independent of the absolute values of conductivity and flake size.

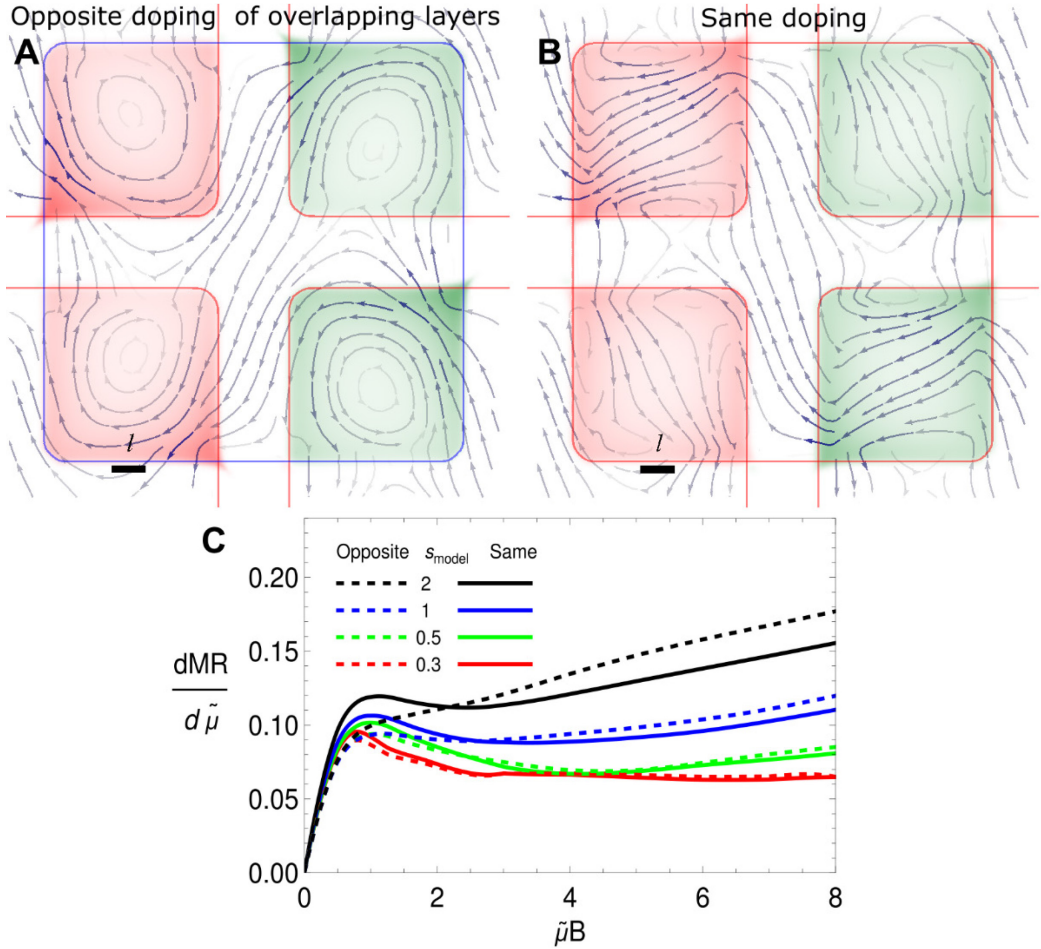

**Figure S2.** Panels A and B: Comparison of current distribution between the opposite (A) and the same (B) type of carriers in the two nanosheets. Plots are made for $s_{model}=1$ and $\tilde{\mu} B = 10$. Panel C: Comparison of the slope of $MR$ between the same/opposite doping combinations.

Hence, in the modeling we can choose the units where $N_{\text{BLG}} \rho_{\text{BLG}} e \, \tilde{\mu} \, \delta E_F = 1$ so that we are left with only two fitting parameters, $s_{model}$ and $\langle E_F \rangle$, and the dimensionless magnetic field, $\tilde{\mu} B$, as a variable. The numerical data for resistance and Hall resistance is taken in the "Hall bar" setup, where we step away from the source and drain contacts to record the voltage. The results for the model consisting of 24 square flakes are further averaged over 15 random realizations of flake Fermi levels.

To aid the numerical stability, the conductivities are further smoothened with a Gaussian kernel, as can be observed in Figure S1, and a small conductivity $10^{-5}\,\sigma_0$ is introduced in the gaps between the flakes.

Finding the parabolic magnetoresistance at low magnetic fields, we can relate the flake mobility, $\tilde{\mu}$, to the effective network "magnetic" mobility, $\mu$, which can be extracted from the experiment, Figure S3.

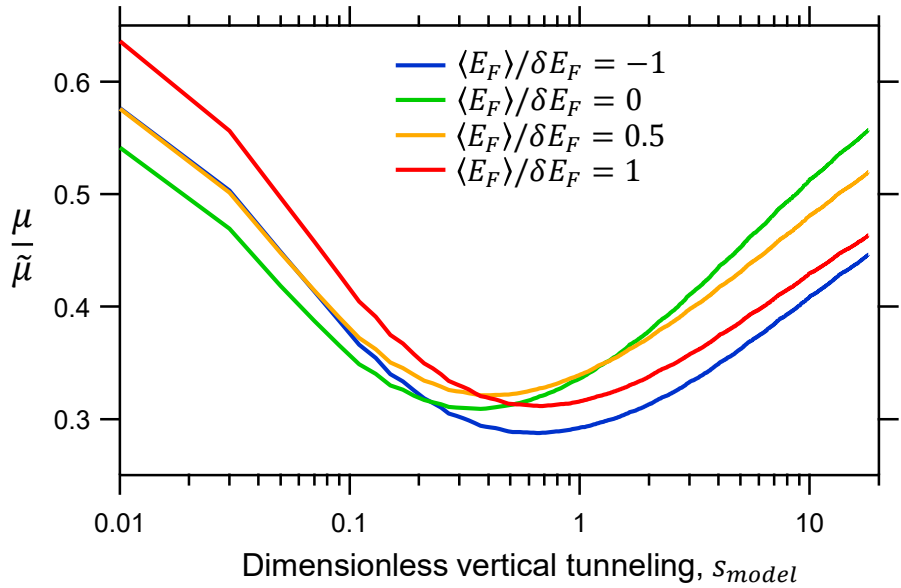

**Figure S3.** The ratio of network mobility to electron mobility of individual flakes, plotted versus dimensionless vertical tunneling.

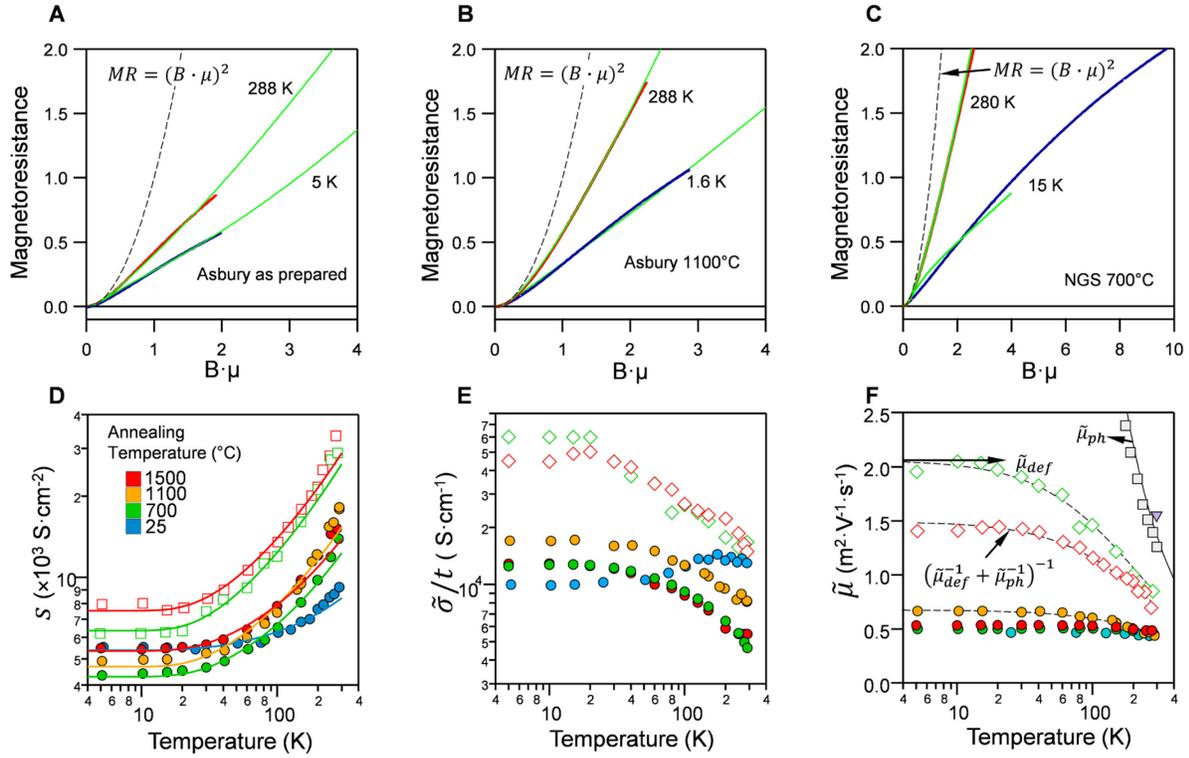

**Figure S4.** Panels A)-C) show the comparison of the numerical model (green lines) with experimental $MR$ data measured at high (red lines) and low temperature (blue lines). The dashed lines show the quadratic dependence of $MR$ on the magnetic field. Note that the magnetic field is normalized to low mobility. Data for three representative samples are shown: (A) as-prepared laminates from graphite supplied by Asbury and two annealed samples from (B) Asbury and (C) NGS graphite. The quality of fit for all samples and temperatures was similar. Panels D)-F) show the temperature dependence of the tunneling interfacial conductance, $s$ in (D), the individual nanosheets' conductivity, $\tilde{\sigma}$, where $t = 4$ nm in (E) and the carrier mobility in (F) for the as-prepared laminates (blue symbols) and the laminates annealed at 700 °C (green symbols), 1100 °C (orange symbols) and 1500 °C (red symbols). Solid curves in panel D) show the theoretical calculations made for the phonon-assisted tunneling model, with the coupling constant $g$ being the only fitting parameter (see Table S1). Filled and empty symbols represent

the data for samples produced from Asbury and NGS graphites. See the main text, Figure 3, for more details.

Finally, we can fit the experimental data for $MR$ and Hall resistance with simulation, finding the best-fit values for $s_{\text{model}}$ and $\langle E_\text{F} \rangle$ for each measured sample at any fixed temperature. Figure 3C of the main text shows examples of the model application. We have also observed that the $MR$ started to show saturation at very large magnetic fields, deviating from the linear trend. This effect appeared in the numerical simulation due to a large but finite resistance in the area between the flakes in the same plane, resulting in the in-plane leakage, which becomes noticeable when the flake resistance is large in high magnetic fields. We suggest that a similar mechanism of in-plane leakage explains the saturation seen in the experimental data for high-quality NGS samples, Figure S4.

As expected, the results of the fitting show that all the annealed samples are strongly inhomogeneous, having an average doping smaller than doping fluctuations, $\langle E_\text{F} \rangle \ll \delta E_\text{F}$, while the as-prepared sample is significantly $p$-doped, $\langle E_\text{F} \rangle \approx -\delta E_\text{F}$, see Table S1.

**S2. Determination of microscopic material properties with the help of the numerical model.**

We use the data on average nanosheet dimensions to relate the numerical modeling results. Since the nanosheets are all different in dimensions and shape, we can only discuss the "*typical*" average values, and the results can be trusted to some numerical prefactor of the order unity. This is valuable for understanding the orders of magnitude and the temperature dependences of in-plane and out-of-plane conductivities. The typical flake conductivity,

$$\tilde{\sigma}_{xx\,\text{flake}} \approx N_{\text{BLG}} \rho_{\text{BLG}} e\, \tilde{\mu}\, \delta E_\text{F} \approx \frac{\sigma_{xx\,\text{phys}}\, t}{\sigma_{xx\,\text{model}}} \tag{S2}$$

is determined by comparing the modeled network conductivity to the experimentally measured one, where we use the average flake thickness $t \approx 4$ nm$^2$. With the help of the density of states in the graphene bilayer $\rho_{BLG} \approx \frac{\gamma_1}{\pi \hbar^2 v_F^2}$ (with $v_F = 10^6$ m s$^{-1}$, $\gamma_1 = 0.35$ eV) and average number of bilayers per flake, $N_{BLG}=3.5$, we may estimate the magnitude of the spread of Fermi levels, $\delta E_F$. The results are presented in Table S1.

**Table S1:** sample parameters estimated from fitting the experimental data with modeling results. The parameters are: $\delta E_F$ – variance of Fermi level distribution; $\langle E_F \rangle$ – average Fermi level of the flakes; $\sigma_{zz,T=0}$ – temperature-independent part of vertical conductivity, ascribed to impurity-assisted tunneling; $g$ – coupling constant of electrons to beating mode of phonons, responsible for temperature-dependent part of vertical conductivity $S$ – Seebeck coefficient measured at 300 K.

| Material/Annealing | $\delta E_F$ (meV) | $\langle E_F \rangle$ (meV) | $\tilde{\sigma}_{zz,T=0}$ (S m$^{-1}$) | $g$ (eV Å$^{-1}$) | $S$ (μV K$^{-1}$) |
|---|---|---|---|---|---|
| Asbury, as-prepared | 80 | -60 | 0.21 | 0.034 | 42 |
| Asbury, 700°C | 30 | -1 | 0.17 | 0.055 | 1.5 |
| Asbury, 1100°C | 55 | 8 | 0.19 | 0.063 | -14.6 |
| Asbury, 1500°C | 30 | -2 | 0.21 | 0.056 | -3.4 |
| NGS, 700°C | 29 | -1 | 0.27 | 0.086 | 0.9 |
| NGS, 1500°C | 70 | -3 | 0.30 | 0.089 | - |

Having found the dimensionless vertical conductivity, $s$, we use the knowledge of the average flake size, $L \approx 5.6$ nm, to estimate the vertical conductivity of randomly oriented flakes,

$$\sigma_{zz} = s\,t = \frac{s_{\text{model}}\, t\, N_{\text{BLG}} \rho_{\text{BLG}} e\, \tilde{\mu}\, \delta E_F}{L^2} = s_{\text{model}} \frac{t^2}{L^2} \frac{\sigma_{xx\,\text{phys}}}{\sigma_{xx\,\text{model}}} \tag{S3}$$

(note that here $\sigma_{xx\,\text{model}}$ is dimensionless since it is measured in the units of $N_{\text{BLG}} \rho_{\text{BLG}} e\, \tilde{\mu}\, \delta E_F$).

## S3. Origins of the linear magnetotransport in graphene laminate

Here we present the physical picture underpinning our numerical results. Before proceeding to magnetoresistance, let us switch off the magnetic field.

To gain understanding, we consider a 1D model of two overlapping flakes with conductivities $\sigma$ and lengths $L$. We are looking at the current transport from point $x = 0$ on flake 1 to point $x = L$ on flake 2. The current conservation equations in the region $[0, L]$ are reduced to

$$j_{1,2} = -\sigma U'_{1,2}\,;\ j_1' = -j_2' = s\,(U_2 - U_1) \tag{S4}$$

with boundary conditions $j_1(L) = j_2(0) = 0$.

The solution features a "*carrier recombination length*", $l = \sqrt{\frac{\sigma}{2s}}$ in the exponent, and the voltage difference between the flakes grows as $U^1 - U^2 \sim e^{\frac{x}{l}}$ near the edge. So, when $L \gg l$, the tunneling happens mainly in the region of width $l$ near the edges of the nanosheet overlap region, while in the opposite limit, when $L \ll l$, the tunneling happens homogeneously through all of the overlap region. The resistance (of a unit length in the "$y$"-direction along the overlap) between points $x = 0$ on flake 1 and $x = L$ on flake 2 behaves as

$$R = \frac{L}{2\sigma} + \frac{l}{\sigma} \coth \frac{L}{2l} \approx \begin{cases} 1/(L\,s), & l \gg L \\ L/(2\,\sigma) + 1/\sqrt{2\,\sigma\,s}, & l \ll L \end{cases} \tag{S5}$$

The first term corresponds to the in-plane resistance, and the second term mostly comes from the tunneling between the flakes. In the weak tunneling regime, $l \gg L$, the second term dominates, and the resistance saturates to $R \approx 1/(L\,s)$, where it corresponds to tunneling being homogeneously distributed over the flake overlap area, in the opposite limit of strong tunneling, $l \ll L$, the tunneling happens only in the region of width $l$ around the boundary of flake overlap region and the resistance behaves as $R \approx L/(2\,\sigma) + 1/\sqrt{2\,\sigma\,s}$. In this regime, the flakes have the same voltage distribution deep inside the overlap region and work together exactly like in a standard two-carrier model. Note that in a wide intermediate range, the $1/\sqrt{s_{\text{model}}}$ term dominates over a constant term.

The magnetoresistance of laminates has two basic regimes: the standard quadratic $MR$ at a low magnetic field, which is very similar but slightly weaker than the $MR$ of a single flake, and a linear regime at a high magnetic field, $\tilde{\mu}B \gg 1$. In this latter regime, the tunneling between the flakes happens mainly along the edges and corners of domains formed by flake intersections, Figure 3B. To understand this behavior, we note that it is the $\tilde{\sigma}_{xx} = \frac{\sigma_0}{1+(\tilde{\mu}B)^2}$ component of conductance that is relevant to the 1D model discussed above, while $\tilde{\sigma}_{xy}$ component describes the current along the edge. Plugging $\sigma = \tilde{\sigma}_{xx}$, we see that with a growing magnetic field, we are inevitably approaching the strong tunneling regime, $l \sim 1/B \ll L$, so the tunneling current between the flakes gets concentrated in the narrow strips of width $l$ along the edge, leading to the linear $MR$ as described in the main text. Here, we also mention that linear $MR$ regime can be achieved even at not-so-high magnetic fields and not-so-weak vertical tunneling, where we only need that $\tilde{\mu}B > 1$, but we can still have $l \sim L$. In this regime, the resistance is determined by the two-terminal magnetoresistance of the flakes, connected via overlapping parts, $R \propto \frac{1}{\sigma_{xy}} \propto \frac{1}{B}$.

Since the resistance at zero magnetic fields behaves approximately as $R(B=0) \propto 1/\sqrt{s}$ in a wide realistic range of $s$, we get for the magnetoresistance:

$$MR(B,s) = \frac{R(B) - R(B=0)}{R(B=0)} \propto B\sqrt{s}, \tag{S6}$$

as shown in Figure S5.

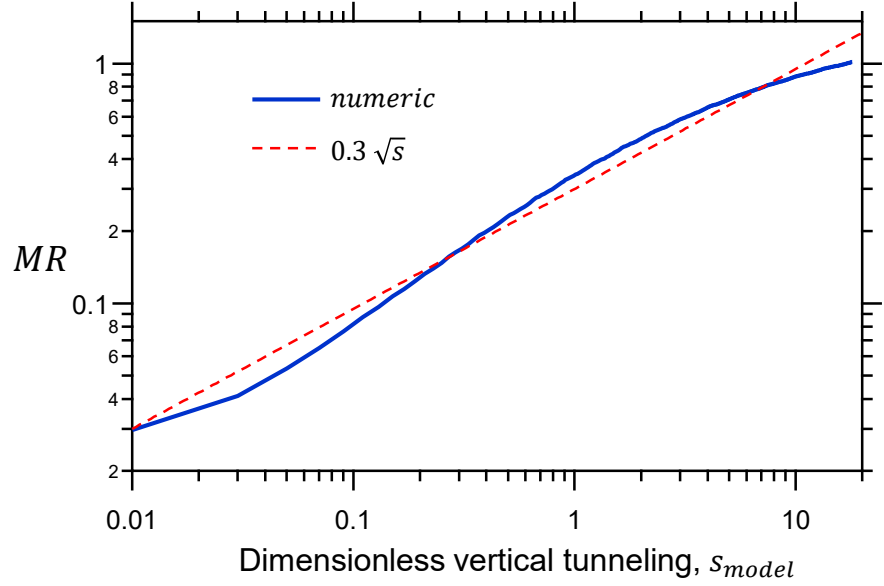

**Figure S5.** The slope of linear $MR$ as a function of vertical tunneling conductance.

## S4. Phonon-assisted tunnelling

This section presents a modified version of considerations in[3] for phonon-assisted tunneling between the flakes. Consider the process of tunneling between layers, assisted by the emission/absorption of the phonons with momentum $q_K \approx 2K \sin\theta/2$, connecting the two twisted Dirac cones, where $K = 4\frac{\pi}{3a} \approx 17$ nm$^{-1}$. The main contribution comes from the flexural breathing modes resulting in the distance beating between adjacent layers[3]. The phonon-mediated current due to bias $V$ can be expressed as

$$I = \frac{4\pi e}{\hbar}\sum_{\{k,k'\}} M_{k'}^{k} f(E_k)\big(1 - f(E_{k'} + eV)\big) - M_{k}^{k'} f(E_{k'} + eV)\big(1 - f(E_k)\big) \tag{S7}$$

and

$$M_{k'}^{k} = |\langle \psi_k | H_{e-ph} \psi_{k'} \rangle|^2 \, [\, n_q \, \delta(E_{k'} - E_k - \hbar\omega_q) + (1+n_q)\delta(E_{k'} - E_k + \hbar\omega_{-q})\,]$$

(S8)

The electron-phonon matrix element may be approximated as

$$|\langle \psi_k | H_{e-ph} | \psi_{k'} \rangle|^2 \approx \frac{1}{4} \frac{g^2 \hbar}{\rho_C \omega(q_K)} \tag{S9}$$

where $g$ – is an electron-phonon coupling constant arising due to modulation of inter-layer hopping of electrons ($g \approx 0.34$ eV Å$^{-1}$ was suggested in[3] for pure twisted graphene, while we expect smaller values due to higher inter-plane distance in stage II intercalated flakes); here $\rho_C$ is the mass density of graphene and the factor ¼ accounts for a probability of an electron in the bilayer to be at the right layer for the tunneling.

Assuming infinitesimal bias V, we then obtain the vertical conductivity

$$s(\theta) = \frac{e^2 g^2 m_{BLG}^2}{4\pi\hbar^3 \rho_C k_B T} \left[\sinh\left(\frac{\hbar\omega_{q_K}}{2 k_B T}\right)\right]^{-2} \tag{S10}$$

The phonon frequency, $\omega_q$, can be estimated via the phonon energy near the $A$-point of graphite as $\omega_q \approx \sqrt{\omega_0^2 + \kappa\, q^4/\rho_C}$ where we expect that the phonon gap is smaller than the graphite value $\hbar\omega_0 < 8.9$ meV (due to larger separation between bilayers) and the bending rigidity $\kappa = 1.65$ eV.[4] We choose $\hbar\omega_0 = 5$ meV, but the result weakly depends on this choice due to dominance of the bending rigidity term for all not-so-small twist angles. A possible internal strain of the flakes would change the phonon dispersion as $\omega_q \approx \sqrt{\omega_0^2 + \kappa\frac{q^4}{\rho_C} + u_{xx} v_L^2 q^2}$, where $u_{xx}$ is strain and $v_L$ is a speed of sound,[5] which leads to a negligible change for any realistic strain values. As the temperature exceeds 200 K, we see an additional contribution to effective tunneling. It might be arising either from the excitation of higher-energy phonon branches[3] leading to an increase of conductivity between the nanosheets or due to increased in-plane conductivity through possible in-plane *p-n* barriers in the flakes, which were not taken into account in our modeling. We do not add this contribution to avoid extra fitting parameters.

To account for the random orientation of flakes in the laminates, we need to map the random twist angles conductivity model to our simpler model, which assumes the same tunneling conductivity between all the flakes. For this, we construct a large 50×50 square lattice electrical network with conductance of each link set according to equation (S1) with a randomly drawn twist angle. Then, we find an effective mean conductance $s_{eff}$ in such a way that the same square lattice but with conductance $s_{eff}$ of each link gives the same network conductivity. It turns out that $s_{eff}$ is rather close to the result for $\theta = 16°$, as shown in Figure S6. We note that this result differs from what one gets by simply averaging $s(\theta)$ over the twist angle.

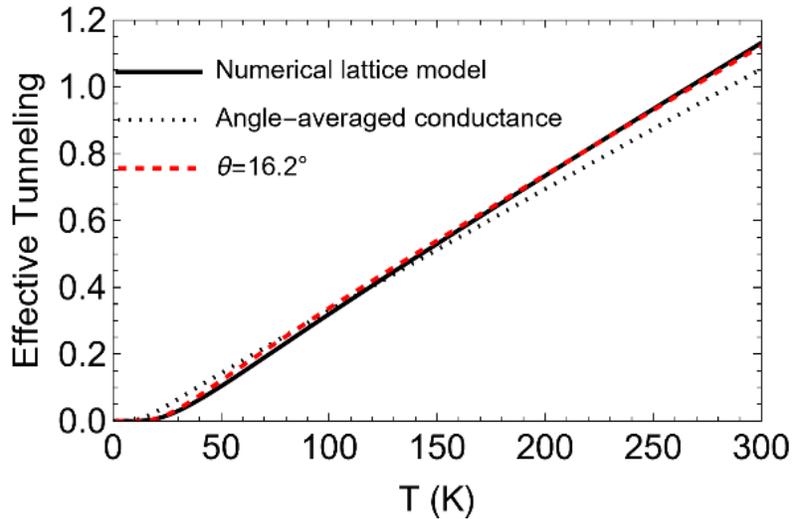

**Figure S6.** Effect of the twist angle of nanosheets on the network conductivity. The conductivity of the random-twist network is well approximated by the conductivity of the network with a fixed twist angle of around 16°.

The result for $\sigma_{zz}$ extracted from comparing the modeling with the experiment is then fitted as $\sigma_{zz} = \sigma_{zz\,impurity} + t\,s_{eff}$, where $s_{eff} \approx s(16°)$ a sum of a temperature-independent value from impurity-assisted tunneling and a phonon-assisted contribution. Here $t \approx 4$ nm is an average thickness of flakes, and the electron-phonon coupling $g$ is treated as a fitting parameter. The results for $\sigma_{zz\,impurity}$ and $g$ fitting parameters are presented in Table S1. Compared to pure twisted graphene bilayers[6] and graphite [7], where $g = 0.34$ eV Å$^{-1}$ [3], we observe a significantly smaller value of electron-phonon coupling $g < 0.1$ eV Å$^{-1}$ and the vertical conductivity, Table S1. This may be due to a notably larger distance between the flakes spaced with pyrene molecules.

**S5. On the magnitude of point defects.**

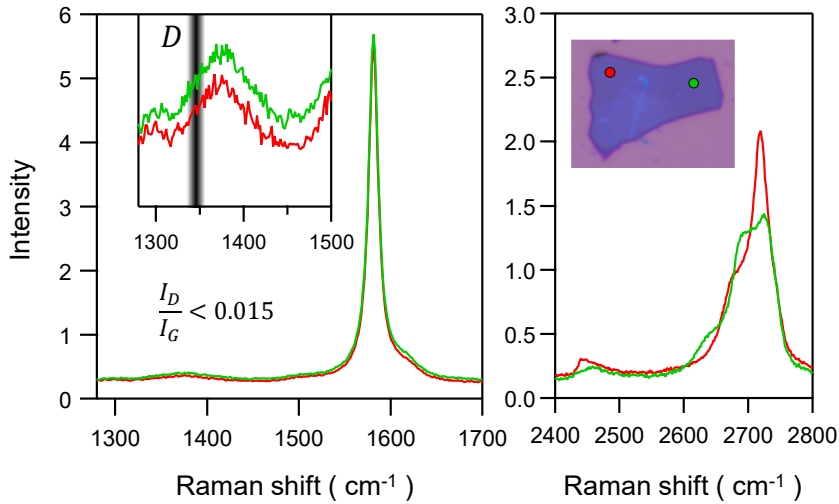

**Figure S7**. Raman spectra captured from a multi-layered graphene nanosheet (shown in the inset of the right panel). The spots where the spectra are captured are shown in a micrograph. The laser energy was 2.33 eV focused on the samples using a ×100 objective. The inset in the left panel shows the spectrum around the D band region. The gray region indicates the Raman shift corresponding to the D band.

The carrier mobility of graphene nanosheets reaches more than 20,000 cm$^2$ V$^{-1}$ s$^{-1}$ at cryogenic temperatures (Figure 3F). This implies that the number density of point defects in graphene nanosheets should be smaller than $\ell_{\text{mfp}}^{-2}$, which is approximately 10$^9$ cm$^{-2}$. Raman spectroscopy has been widely used to quantify the amount of such defects in graphene and graphite, often by studying the ratio of the D to G bands.[8]

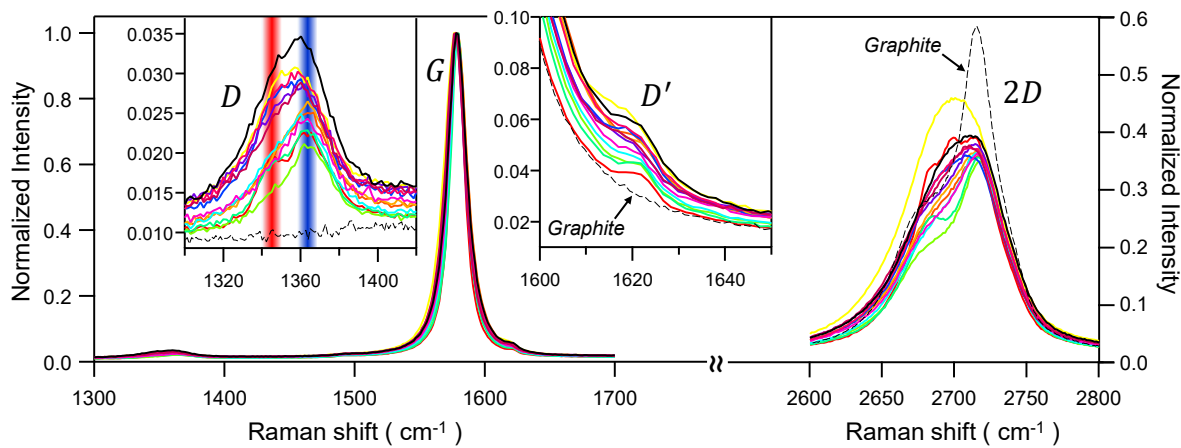

**Figure S8.** Raman spectra captured from 13 different spots of a graphene laminate annealed at 1500 °C made from NGS graphite crystals. The dashed line shows the graphite spectrum. The D band often showed an asymmetric feature, a characteristic feature of intact graphite and graphene edges.[9, 10]

Figure S7 shows the Raman spectrum captured from nanosheets made from high-quality graphite. As discussed in our previous work,[2] the presence of sulfonated pyrene (s-Py) in the nanosheets interferes with the D band signal from possible point defects in the sample. A weak, broad peak is observed at around 1375 cm$^{-1}$, originating from s-Py molecules.[2] The absence of any pronounced band at 1340 cm$^{-1}$, where the D peak originated from point-like defects, is expected, indicating a small number density of point-like defects. Based on the intensity of the spectrum around this energy, we expect $\frac{I_D}{I_G}$ to be smaller than 0.015, which is equivalent to defect density smaller than $3.4 \pm 0.9 \times 10^9$ cm$^{-2}$.[8]

The Raman peaks due to s-Py disappeared after annealing the laminates above nearly 1000 °C. We often detected weak D bands in the Raman spectrum captured from annealed laminates. However, it originates from edge defects rather than point-like defects. Given that the optical penetration depth for graphite (~60 nm at 532 nm) is much larger than the nanosheet thickness (~4 nm), it is likely that the spectrum is not from the top layer but a few nanosheets combined. In the case of laminates made from NGS graphite (Figure S8), we often observed an asymmetric D band ascribed to the edge defects of graphene or graphite.[9, 10] In addition, D` band around 1620 cm$^{-1}$ was always observed (Figure S8), and the intensity ratio of D to D` band was also consistent with the edge defect type.[11] We, therefore, conclude that the point-like defect is negligible in our samples, which is consistent with the high mobility of nanosheets at cryogenic temperatures.

## S8. Elemental composition of the laminates:

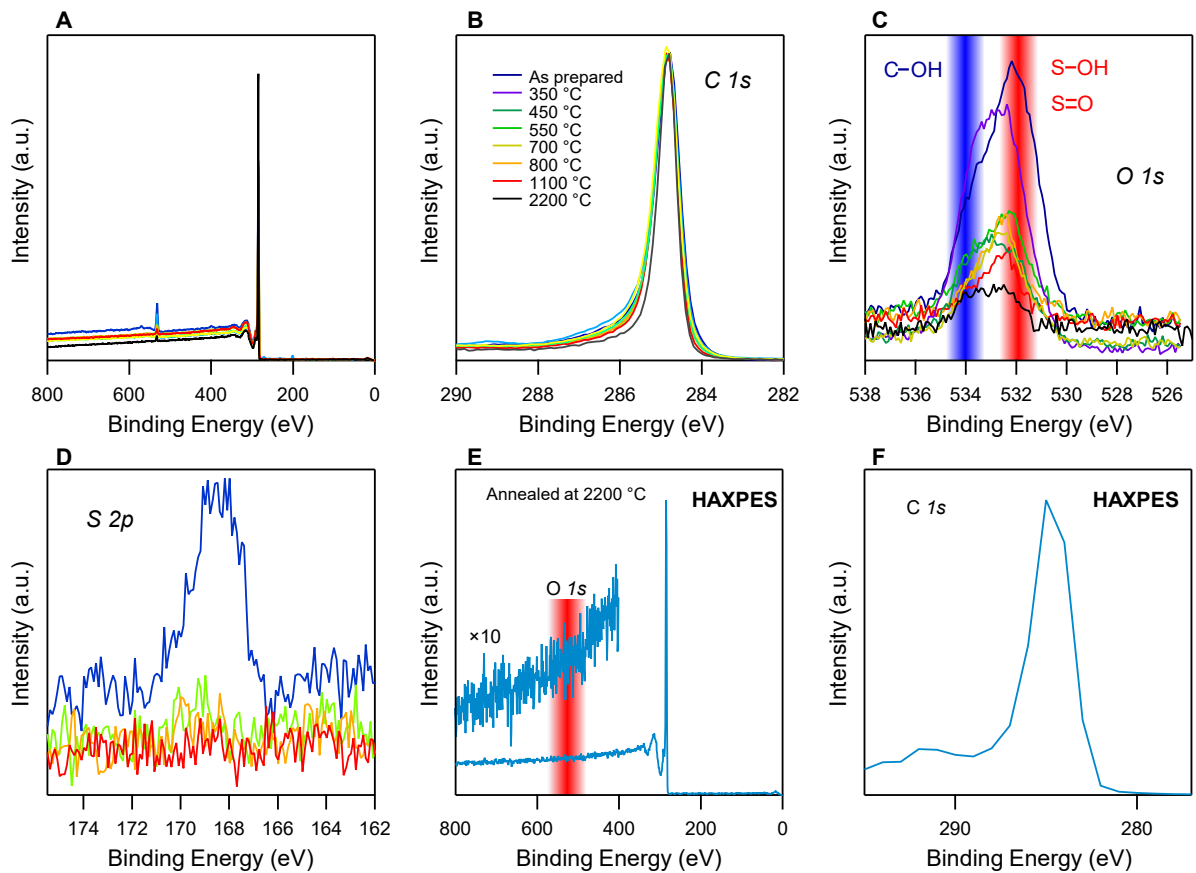

**Figure S9.** Elemental analysis of graphene laminates. Panels (**A-D**) show the XPS spectra of laminates annealed at different temperatures. Panels (**E**) and (**F**) show the HAXPES spectrum of the laminate annealed at 2200 °C. The oxygen content was below the detection limit. The C 1s peak shows a distinct graphitic structure.

XPS analysis showed that the as-prepared sample contains 0.9, 6.3 and 92.8 atomic percent sulfur, oxygen and carbon, respectively (Figure S9). The sulfur signal is most likely from tetra-sulfonated pyrene ($C_{16}H_{10}O_{12}S_4$). Upon annealing, most of the sulfur evaporated through de-sulfonating of s-Py. Indeed, this is consistent with the disappearance of the sulfur peak in the XPS spectra of annealed samples (Figure S9D).

As every sulfonated group contains one sulfur and three oxygen atoms, reduction in oxygen content is also consistent with de-sulfonating s-Py upon annealing (Figure S9C). XPS analysis of laminates annealed at 2200 °C showed 0.92 atomic percent oxygen. We ran hard X-ray photoelectron spectroscopy (HAXPES) on the same sample to rule out the oxygen associated with the surface. The oxygen content was below the detection limit (Figures S9E and S9F). This corresponds to a carbon content of 100.00%. This is consistent with the high carrier mobility in our samples.

**References**

(1) Zou, K.; Hong, X.; Zhu, J. Effective mass of electrons and holes in bilayer graphene: Electron-hole asymmetry and electron-electron interaction. *Physical Review B* **2011**, *84* (8), 085408.

(2) Moazzami Gudarzi, M.; Asaad, M.; Mao, B.; Pinter, G.; Guo, J.; Smith, M.; Zhong, X.; Georgiou, T.; Gorbachev, R.; Haigh, S. J.; et al. Chlorosulfuric acid-assisted production of functional 2D materials. *npj 2D Materials and Applications* **2021**, *5* (1), 35. DOI: 10.1038/s41699-021-00215-2.

(3) Perebeinos, V.; Tersoff, J.; Avouris, P. Phonon-Mediated Interlayer Conductance in Twisted Graphene Bilayers. *Physical Review Letters* **2012**, *109* (23), 236604. DOI: 10.1103/PhysRevLett.109.236604.

(4) Nicklow, R.; Wakabayashi, N.; Smith, H. G. Lattice Dynamics of Pyrolytic Graphite. *Physical Review B* **1972**, *5* (12), 4951-4962. DOI: 10.1103/PhysRevB.5.4951.

(5) Ochoa, H.; Castro, E. V.; Katsnelson, M.; Guinea, F. Temperature-dependent resistivity in bilayer graphene due to flexural phonons. *Physical Review B* **2011**, *83* (23), 235416.

(6) Kim, Y.; Yun, H.; Nam, S.-G.; Son, M.; Lee, D. S.; Kim, D. C.; Seo, S.; Choi, H. C.; Lee, H.-J.; Lee, S. W.; et al. Breakdown of the Interlayer Coherence in Twisted Bilayer Graphene. *Physical Review Letters* **2013**, *110* (9), 096602. DOI: 10.1103/PhysRevLett.110.096602.

(7) Koren, E.; Leven, I.; Lörtscher, E.; Knoll, A.; Hod, O.; Duerig, U. Coherent commensurate electronic states at the interface between misoriented graphene layers. *Nature Nanotechnology* **2016**, *11* (9), 752-757. DOI: 10.1038/nnano.2016.85.

(8) Cançado, L. G.; Jorio, A.; Ferreira, E. H. M.; Stavale, F.; Achete, C. A.; Capaz, R. B.; Moutinho, M. V. O.; Lombardo, A.; Kulmala, T. S.; Ferrari, A. C. Quantifying Defects in Graphene via Raman Spectroscopy at Different Excitation Energies. *Nano Letters* **2011**, *11* (8), 3190-3196. DOI: 10.1021/nl201432g.

(9) Cançado, L. G.; Pimenta, M. A.; Neves, B. R. A.; Dantas, M. S. S.; Jorio, A. Influence of the Atomic Structure on the Raman Spectra of Graphite Edges. *Physical Review Letters* **2004**, *93* (24), 247401. DOI: 10.1103/PhysRevLett.93.247401.

(10) Li, Q.-Q.; Zhang, X.; Han, W.-P.; Lu, Y.; Shi, W.; Wu, J.-B.; Tan, P.-H. Raman spectroscopy at the edges of multilayer graphene. *Carbon* **2015**, *85*, 221-224.

(11) Eckmann, A.; Felten, A.; Mishchenko, A.; Britnell, L.; Krupke, R.; Novoselov, K. S.; Casiraghi, C. Probing the Nature of Defects in Graphene by Raman Spectroscopy. *Nano Letters* **2012**, *12* (8), 3925-3930. DOI: 10.1021/nl300901a.